\documentclass[submission,copyright,creativecommons]{eptcs}
\providecommand{\event}{GASCom 2026} 

\usepackage{iftex}
\usepackage{amsmath}
\usepackage{amssymb}
\usepackage{mleftright}
\usepackage{amsthm}
\usepackage{graphicx}
\usepackage{tikz}
\usetikzlibrary{decorations.pathmorphing,arrows.meta}
\usetikzlibrary{decorations.pathreplacing}
\newtheorem{theorem}{Theorem}
\newtheorem{lemma}{Lemma}

\ifpdf
  \usepackage{underscore}         
  \usepackage[T1]{fontenc}        
\else
  \usepackage{breakurl}           
\fi

\title{Random Generation of $k$-coloured Motzkin Paths}
\author{Elena Barcucci
\institute{University of Florence\\ Italy}
\email{elena.barcucci@unifi.it}
\and
Antonio Bernini
\institute{University of Florence\\ Italy}
\email{antonio.bernini@unifi.it}
\and
Stefano Bilotta
\institute{University of Florence\\ Italy}
\email{stefano.bilotta@unifi.it}
\and
Renzo Pinzani
\institute{University of Florence\\ Italy}
\email{renzo.pinzani@unifi.it}
}
\def\titlerunning{Random Generation of $k$-coloured Motzkin Paths}
\def\authorrunning{E. Barcucci, A. Bernini, S. Bilotta \& R. Pinzani}
\begin{document}

\maketitle

\begin{abstract}
We study $k$-coloured Motzkin paths, namely Motzkin paths in which
horizontal steps can be coloured in $k$ different ways, and investigate
their connection with the number of prefixes ending at odd height
from both an analytical and a combinatorial point of view.
Moreover, the combinatorial approach provides a random
generation algorithm for $k$-coloured Motzkin paths in linear-time.
\end{abstract}

\section{Foreword}
The research presented in this contribution was initiated in early 2003 by two of the present authors, Elena Barcucci and Renzo Pinzani, together with Alberto Del Lungo, a past student of theirs, at that time full professor at the University of Siena. He suddenly died on June 3rd
2003, when he was only 38 years old. The pain of his death for the two was so high that they completely forgot the manuscript, describing some results about the mentioned research, which Alberto gave them a few days before his death. During last summer, Elena, while recovering the documents in her room because of her retirement, found the manuscript. As a consequence, Elena and Renzo asked the co-authors Antonio Bernini and Stefano Bilotta, two other past students of theirs, for help in revising and completing the results. They accepted with enthusiasm and so the work was completed.

\section{Introduction}
A \emph{Motzkin path} is a lattice path in the plane that starts at $(0,0)$, ends on the $x$-axis, never goes below it, and consists of three types of steps: an up step $(1,1)$, a down step $(1,-1)$, and a horizontal step $(1,0)$. Motzkin paths are classical objects in enumerative combinatorics enumerated by the famous Motzkin numbers, originally introduced by T.~Motzkin \cite{Motzkin}. They have been extensively investigated due to their
connections with Catalan structures, continued fractions, and
various combinatorial sequences \cite{GKP,Stanley1}. Several
extensions have been considered, including coloured Motzkin paths
and weighted variations, which enrich the combinatorial structure
while preserving recursive decompositions \cite{Deutsch}. In addition, we also recall their role in the generation of cross bifix-free words \cite{BBPPS}, together with the study of their generating function under pattern-avoidance constraints \cite{BBCF}.

Many efficient algorithms to randomly generate Motzkin words of fixed
size are known. Several of them are based on rejection methods, as in \cite{Alo2023}, while  Barcucci et al.\ \cite{BPS94, BPS95} proposed
an efficient algorithm based on an anticipated rejection method, known as the Florentine algorithm \cite{Bacher}. 

In this paper we consider
$k$-coloured Motzkin paths, namely Motzkin paths in which horizontal
steps can be coloured in $k$ different ways, while rise and fall steps
remain uncoloured.
Let $m_n$ denote the number of $k$-coloured Motzkin paths of length $n$.
We are interested in the enumeration of left factors (prefixes) of such
paths according to the parity of their final height.
More precisely, let $f_n^{(o)}$ (resp.\ $f_n^{(e)}$) be the number of
prefixes of length $n$ ending at odd (resp.\ even) height, and let
$f_n$ be the total number of prefixes of length $n$.

In this paper we consider a relation between
$k$-coloured Motzkin paths and prefixes ending at odd height, namely
\[
f_{n+1}^{(o)} = (n+1)m_n,
\qquad \text{for } n \ge 0.
\]
We provide two different proofs of this identity.
The first one is analytic and is based on the functional equation
satisfied by the generating function of $k$-coloured Motzkin paths.
The second one is purely combinatorial and relies on a bijective
transformation together with Raney's lemma \cite{R} on cyclic shifts of
integer sequences.

Besides giving two independent proofs of the same enumerative
identity, the combinatorial approach also leads to a random
generation algorithm for $k$-coloured Motzkin paths and
a linear-time complexity analysis of it.

%
\section{Enumeration of Motzkin prefixes}
We give two different proofs of the following
\bigskip
\begin{theorem}\label{teo}
	For $n \geq 0$,
	\[
	f_{n+1}^{(o)} = (n+1) m_n.
	\]
\end{theorem}

\subsection*{\small{Analytical proof} (sketch)}

It is not difficult to verify that 

\[
f_{n+1}^{(e)} = 2 f_n^{(o)} + k f_n^{(e)},
\]
and 
\[
f_{n+1}^{(o)} = k f_n^{(o)} + 2 f_n^{(e)} - m_n .
\]

\noindent with initial conditions
$f_0^{(e)} = 1, \ f_0^{(o)} = 0, \
f_1^{(e)} = k, \ f_1^{(o)} = 1.$

\smallskip

We now define the generating functions

\[
E(t) = \sum_{n \ge 0} f_n^{(e)} t^n,
\qquad
O(t) = \sum_{n \ge 0} f_n^{(o)} t^n,
\qquad
M(t) = \sum_{n \ge 0} m_n t^n.
\]
From the above recurrence relations we obtain
\[
\begin{cases}
	E(t) = 2t\, O(t) + k t\, E(t) + 1,\\
	O(t) = k t\, O(t) + 2t\, E(t) - t\, M(t),
\end{cases}
\]
whose solution is given by
\[
E(t) = \frac{2t O(t) + 1}{1 - k t},
\]
and
\begin{equation}\label{odt}
	O(t) =
	\frac{(1 - kt)t M(t) - 2t}
	{4t^2 - (1 - kt)^2}.
\end{equation}
Moreover, we recall the functional equation satisfied by $M(t)$:

\begin{equation}\label{mdt}
	M(t) = 1 + k t M(t) + t^2 M(t)^2 .
\end{equation}
Setting
$
\overline{M}(t) = t M(t)
$
and performing some straightforward manipulations, we obtain
\begin{equation*}
	t \frac{d}{dt}\overline{M}(t)
	=
	\frac{\overline{M}(t)}
	{1 - kt - 2t\overline{M}(t)}.
\end{equation*}
We have $f_{n+1}^{(o)} = (n+1) m_n$ if and only if $O(t)=	t \frac{d}{dt}\overline{M}(t)$. This latter identity follows from equations \eqref{odt} and \eqref{mdt}.

\subsection*{\small{Combinatorial proof} (sketch)}

Let $P$ be a $k$-coloured Motzkin factor whose last point has odd height
$h=2i+1, \text{for } i \geq 1$. Then $P$ can be factorized as
$$
P=P_1U_1P_2U_2\ldots P_{2i+1}U_{2i+1}P_{2i+2}
$$
where $P_{\ell}$ is a $k$-coloured Motzkin path (possibly the empty path) and $U_{\ell}$ is a rise step. This decomposition highlights the
rise steps that actually contribute to the final height of $P$.

We perform a transformation $\Gamma$ (based on a classical reflection argument) by turning the first $i+1$ rise steps $U_{\ell}$ into $i+1$ fall steps, denoted by $D_{\ell}$, obtaining the path $P'$:
$$
P'=P_1D_1P_2D_2\ldots P_{i+1}D_{i+1}P_{i+2}U_{i+2}\ldots P_{2i+1}U_{2i+1}P_{2i+2}.
$$
The following figure shows an example of the action of $\Gamma$ on a $2$-coloured Motzkin prefix of final height $5$ (so that $i=2$), together with its explicit decomposition (note that in the figure $P'$ is weakly below the $x$-axis, but in general this is not the case).
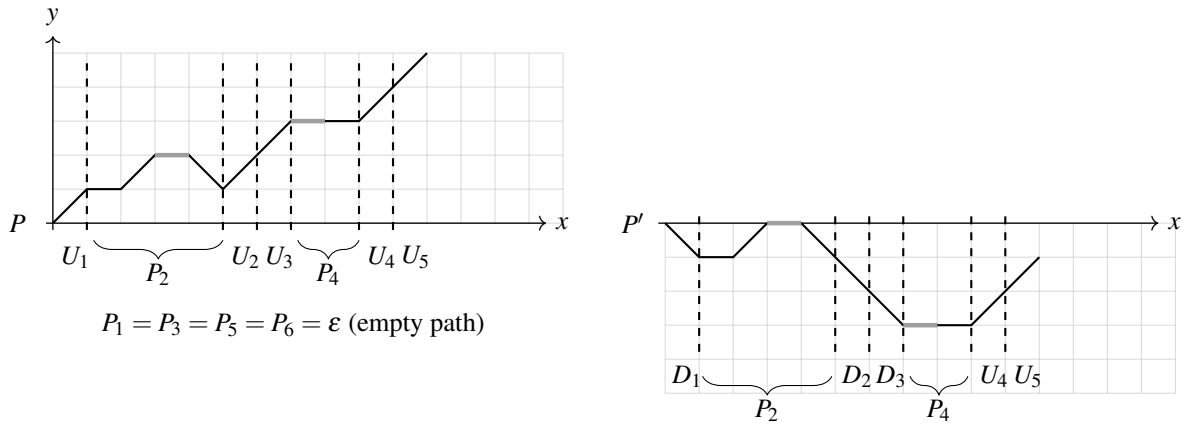
\begin{center}
\begin{tikzpicture}[scale=0.45, every node/.style={font=\small}]
	\begin{scope}
		\draw[step=1cm, gray!30, very thin] (0,0) grid (15,5);
		\draw[->, thick,line width=0.5pt] (-0.2,0) -- (14.5,0) node[right] {$x$};
		\draw[->, thick,line width=0.5pt] (0,-0.2) -- (0,5.5) node[above] {$y$};
		\draw[thick] (0,0) -- (1,1) -- (2,1) -- (3,2) -- (4,2) -- (5,1) -- (6,2) -- (7,3) -- (8,3) -- (9,3) -- (10,4) -- (11,5);
		\draw[thick,gray!75,line width=1.7pt] (3,2) -- (4,2) (7,3) -- (8,3);
		\node[left] at (-0.5,0) {$P$};
		\node[left] at (1.4,-1) {$U_1$};
        \node[left] at (6.4,-1) {$U_2$};
        \node[left] at (7.4,-1) {$U_3$};
        \node[left] at (10.4,-1) {$U_4$};
        \node[left] at (11.4,-1) {$U_5$};
        \node[left] at (13,-3) {$P_1=P_3=P_5=P_6=\epsilon$ (empty path)};
        \draw[dashed, line width=0.8pt] (1,-0.2) -- (1,4.7);
        \draw[dashed, line width=0.8pt] (5,-0.2) -- (5,4.7);
        \draw[dashed, line width=0.8pt] (6,-0.2) -- (6,4.7);
        \draw[dashed, line width=0.8pt] (7,-0.2) -- (7,4.7);
        \draw[dashed, line width=0.8pt] (9,-0.2) -- (9,4.7);
        \draw[dashed, line width=0.8pt] (10,-0.2) -- (10,4.7);
\draw[decorate, decoration={brace, amplitude=7pt, mirror}] (1.2,-0.5) -- (5,-0.5)  node[midway, below=5pt] {$P_2$};

\draw[decorate, decoration={brace, amplitude=7pt, mirror}] (7.2,-0.5) -- (9,-0.5)  node[midway, below=5pt] {$P_4$};
	\end{scope}
	
	\begin{scope}[xshift=18cm]
		\draw[step=1cm, gray!30, very thin] (0,0) grid (15,-5);
		\draw[->, thick,line width=0.5pt] (-0.2,0) -- (14.5,0) node[right] {$x$};
		\draw[thick] (0,0) -- (1,-1) -- (2,-1) -- (3,0) -- (4,0) -- (5,-1) -- (6,-2) -- (7,-3) -- (8,-3) -- (9,-3) -- (10,-2) -- (11,-1);
		\draw[thick,gray!75,line width=1.7pt] (3,0) -- (4,0) (7,-3) -- (8,-3);
		\node[left] at (-0.3,0) {$P'$};
		
		\node[left] at (1.4,-4.5) {$D_1$};
		\node[left] at (6.4,-4.5) {$D_2$};
		\node[left] at (7.4,-4.5) {$D_3$};
		\node[left] at (10.4,-4.5) {$U_4$};
		\node[left] at (11.4,-4.5) {$U_5$};
		\draw[dashed, line width=0.8pt] (1,0.2) -- (1,-3.8);
		\draw[dashed, line width=0.8pt] (5,0.2) -- (5,-3.8);
		\draw[dashed, line width=0.8pt] (6,0.2) -- (6,-3.8);
		\draw[dashed, line width=0.8pt] (7,0.2) -- (7,-3.8);
		\draw[dashed, line width=0.8pt] (9,0.2) -- (9,-3.8);
		\draw[dashed, line width=0.8pt] (10,0.2) -- (10,-3.8);
		\draw[decorate, decoration={brace, amplitude=7pt, mirror}] (1.1,-4.5) -- (4.9,-4.5)  node[midway, below=5pt] {$P_2$};
		
		\draw[decorate, decoration={brace, amplitude=7pt, mirror}] (7.2,-4.5) -- (8.9,-4.5)  node[midway, below=5pt] {$P_4$};
		
	\end{scope}
\end{tikzpicture}
\end{center}

One can prove that this transformation defines a bijection
between the set of $k$-coloured Motzkin prefixes of length~$n$ having odd 
final height and the set of paths of length~$n$ ending at ordinate~$-1$. 

We encode rise, fall, and horizontal steps (independently of their colour) 
by the integers $1$, $-1$, and $0$, respectively. Each path can then be read as a sequence of integers, and the one corresponding
to $P'$ clearly has sum equal to $-1$. From the Raney's Lemma \cite{R} (which can be found also in \cite[Chapter 7, p. 360]{GKP}) it is possible to deduce the following:

\begin{lemma}\label{Raney}
	If $(x_1,\dots,x_{n+1})$ is a sequence of integers whose sum is $-1$,
	exactly one of its cyclic shifts	
	\[
	(x_2,x_3,\dots,x_{n+1},x_1),(x_3,x_4,\dots,x_{n+1},x_1.x_2),\dots,
	(x_{n+1},x_1,\dots,x_{n-1,}x_n)
	\]
has all of its partial sums up to the $n$-th entry non-negative.
\end{lemma}

\noindent
From the above Lemma there exists an unique cyclic shift of the encoding of $P'$ which is a $k$-coloured Motzkin path except that for the last integer.
Moreover, the identification of this unique cyclic shift can be performed in linear time: it suffices to locate the index immediately following the position where the first minimum of the partial sums is attained.

 Since $\Gamma$ is a bijection and each cyclic shift has final height equal to $-1$, then each of the $n+1$ cyclic shifts has a different corresponding $k$-coloured Motzkin prefix with odd final height. Therefore: 

\[
f_{n+1}^{(0)} = (n+1) m_n.
\]
Continuing the example shown above, the encoding of $P$ is $(1,0,1,0,-1,1,1,0,0,1,1)$ and the encoding of $P'$ is $(-1,0,1,0,-1,-1,-1,0,0,1,1)$. Its unique cycle shift corresponding to the $k$-coloured Motzkin path is $(0,0,1,1,-1,0,1,0,-1,-1,-1)$ illustrated below:  

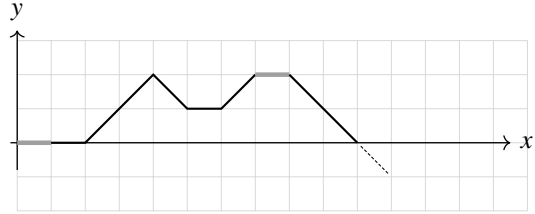
\begin{center}
\begin{tikzpicture}[scale=0.45, every node/.style={font=\small}]
		\draw[step=1cm, gray!30, very thin] (0,0) grid (15,5);
		\draw[->, thick, line width=0.5pt] (-0.2,2) -- (14.5,2) node[right] {$x$};
		\draw[->, thick,line width=0.5pt] (0,1.2) -- (0,5.3) node[above] {$y$};
		\draw[thick,gray!75,line width=1.7pt] (0,2) -- (1,2);
		\draw[thick] (1,2) -- (2,2) -- (3,3) -- (4,4) -- (5,3)--(6,3)--(7,4);
		\draw[thick,gray!75,line width=1.7pt] (7,4) -- (8,4);
		\draw[thick] (8,4) -- (9,3) -- (10,2);
		\draw[thin, dash pattern=on 1.2pt off 0.7pt] (10.1,1.9) -- (10.9,1.1);
\end{tikzpicture}
\end{center}

Note that, each other cyclic shift different from $\mathcal S$ 
corresponds to a $k$-coloured Motzkin prefix different from $P$.

\section{The random generation algorithm}

From the previous combinatorial proof we obtain an algorithm to randomly generate
a $k$-coloured Motzkin path:

\begin{enumerate}
	\item use the Florentine algorithm \cite{BPS94} to 
	randomly generate a $k$-coloured Motzkin prefix $F$ of length $n+1$;
	\item while $F$ has even final height, return to step $1$;
	\item perform the $\Gamma$ transformation on $F$;
	\item perform the cyclic shifts of Lemma \ref{Raney} and identify the unique Motzkin path of length $n$ associated to the $n+1$ cyclic shifts of length $n+1$.
\end{enumerate}

\subsection{The complexity}

The expected number of extracted characters (or calls to \emph{random})
for generating a $k$-coloured Motzkin left factor of length $n$ is \cite{BPS94}

\[
2n - \frac{1}{2}\sqrt{(k+2)\pi (n+1)}
+ \frac{6+k}{4} + O(1).
\]
Therefore, the first step can be performed in linear time.
For the second step, establishing that the final height of the newly generated prefix $F$ is clearly performed in constant time. Nevertheless, the generation of $F$ having final odd height depends on the ratio 
$\frac{f_n^{(0)}}{f_n}.$
	
	From Theorem \ref{teo} we have 
	$\frac{f_n^{(0)}}{f_n}
	=
	\frac{n m_{n-1}}{f_n},
	$
	and from the recursive construction of $k$-coloured Motzkin left factors we have $m_n = (k+2)f_n - f_{n+1}$.
	Thus,	
	\[
	\frac{f_n^{(0)}}{f_n}
	=
	\frac{n\big((k+2)f_{n-1} - f_n\big)}{f_n}.
	\]
	Being 
	$\displaystyle f(t)=\sum_{n\geq0} f_n t^n=\frac{1}{2t}\left( \sqrt{\frac{1+t}{1-3t}}-1\right)$, see for example \cite{BPS94}, it is easy to prove (via the Darboux's method \cite{D}) that	
	\[
	f_n \sim
	\frac{1}{\sqrt{k+2}}
	\frac{(k+2)^n}{\sqrt{\pi n}}.
	\]	
	Therefore, 	
	\[
	\frac{f_n^{(0)}}{f_n}
	\sim \ldots \sim \frac{n}{\sqrt{n-1}} \left(\frac{1}{\sqrt{n}+\sqrt{n-1}}\right)
	\sim
	\frac{1}{2}.
	\]
In the following table, we show some numerical results for $k=1$:

\[
\begin{array}{r|l}
	n & f_n^{(0)}/{f_n} \\ \hline
	1 & 0.5 \\
	2 & 0.5 \\
	3 & 0.4653 \\
	4 & 0.4574 \\
	5 & 0.46875 \\
	6 & 0.47195 \\
	7 & 0.47600 \\
	8 & 0.47856 \\
	9 & 0.48081 \\
	10 & 0.48257 \\
	20 & 0.49096 \\
	30 & 0.49310 \\
	40 & 0.49539 \\
	50 & 0.49630
\end{array}
\]

Therefore, the average computational cost of the first and second steps grows asymptotically as $4n$. The third and fourth steps can be performed in linear time. Then, the average complexity of the algorithm is linear in $n$.

\section{Conclusions}
The results herein contained are surely not up to date. Indeed better performing algorithms do exist in order to generate Motzkin prefixes \cite{Bacher} which could improve step 1 of our algorithm. Nevertheless the authors believe that these results give a new contribution to the field of the random generation of combinatorial objects.

\normalsize
\bibliographystyle{eptcs}
\bibliography{generic}
\end{document}